\documentclass[conference]{IEEEtran}
\IEEEoverridecommandlockouts
\usepackage{cite}
\usepackage{amsmath,amssymb,amsfonts}
\usepackage{algorithmic}
\usepackage{graphicx}
\usepackage{textcomp}
\usepackage{authblk}
\usepackage{xcolor}
\def\BibTeX{{\rm B\kern-.05em{\sc i\kern-.025em b}\kern-.08em
    T\kern-.1667em\lower.7ex\hbox{E}\kern-.125emX}}

\begin{document}

\title{Optimized Memory System Architecture for\\VESA VDC-M Decoder with Multi-Slice Support}

\author[1]{Hannah Yang}
\author[1]{Sohyeon Kim}
\author[2]{Saeyeon Kim}
\author[2]{Jiyoung Lee}
\author[2]{Huijin Roh}
\author[1]{Ji-Hoon Kim}
\affil[1]{\small{Department of Electronic Engineering, Hanyang University, Seoul, Republic of Korea}}
\affil[2]{\small{Department of Electronic and Electrical Engineering, Ewha Womans University, Seoul, Republic of Korea}}
\affil[ ]{\small{E-mail: hannahyang714@gmail.com}}

\maketitle

\begin{abstract}
Video compression plays a pivotal role in managing and transmitting large-scale display data, particularly given the growing demand for higher resolutions and improved video quality. This paper proposes an optimized memory system architecture for Video Electronics Standards Association (VESA) Display Compression-M (VDC-M) decoder, characterized by its substantial on-chip buffer requirements. We design and analyze three architectures categorized by optimization levels and management complexity. Our strategy focuses on enhancing line buffer access scheduling and minimizing reconstruction buffer, targeting prediction and multi-slice operation that are the major resource consumers in the decoder. By adjusting line delay and segmenting SRAM bank alongside reconstructed block forwarding, we achieve a 33.3\% size reduction in the line buffer and 77.3\% in the reconstruction buffer compared to Baseline VDC-M decoder. Synthesized using a 28 nm CMOS process, the proposed architecture achieve a 31.5\% reduction in gate count of the decoder backend hardware, supporting real-time performance with up to 96.45 fps for 4K UHD resolution at 200 MHz operating frequency and a throughput of 4 pixels per cycle.
\end{abstract}


\section{Introduction}
Rapid expansion of display technologies has been accompanied by continuous improvements in resolution, frame rates, and pixel depth, requiring increased display bandwidth \cite{b1}. Over the past decade, the bandwidth requirements for high-end mobile devices have increased 23x, yet the capacity of the physical interfaces (PHY) has improved only half of that \cite{b2}. Video compression is widely applied to bridge this gap. The Video Electronics Standards Association (VESA) introduced the VESA Display Compression-M (VDC-M), a successor to the previously established Display Stream Compression (DSC) standard, as a cost-effective, low-latency, and visually lossless video stream codec.

\begin{figure}[t]
\centerline{\includegraphics[width=8.1cm]{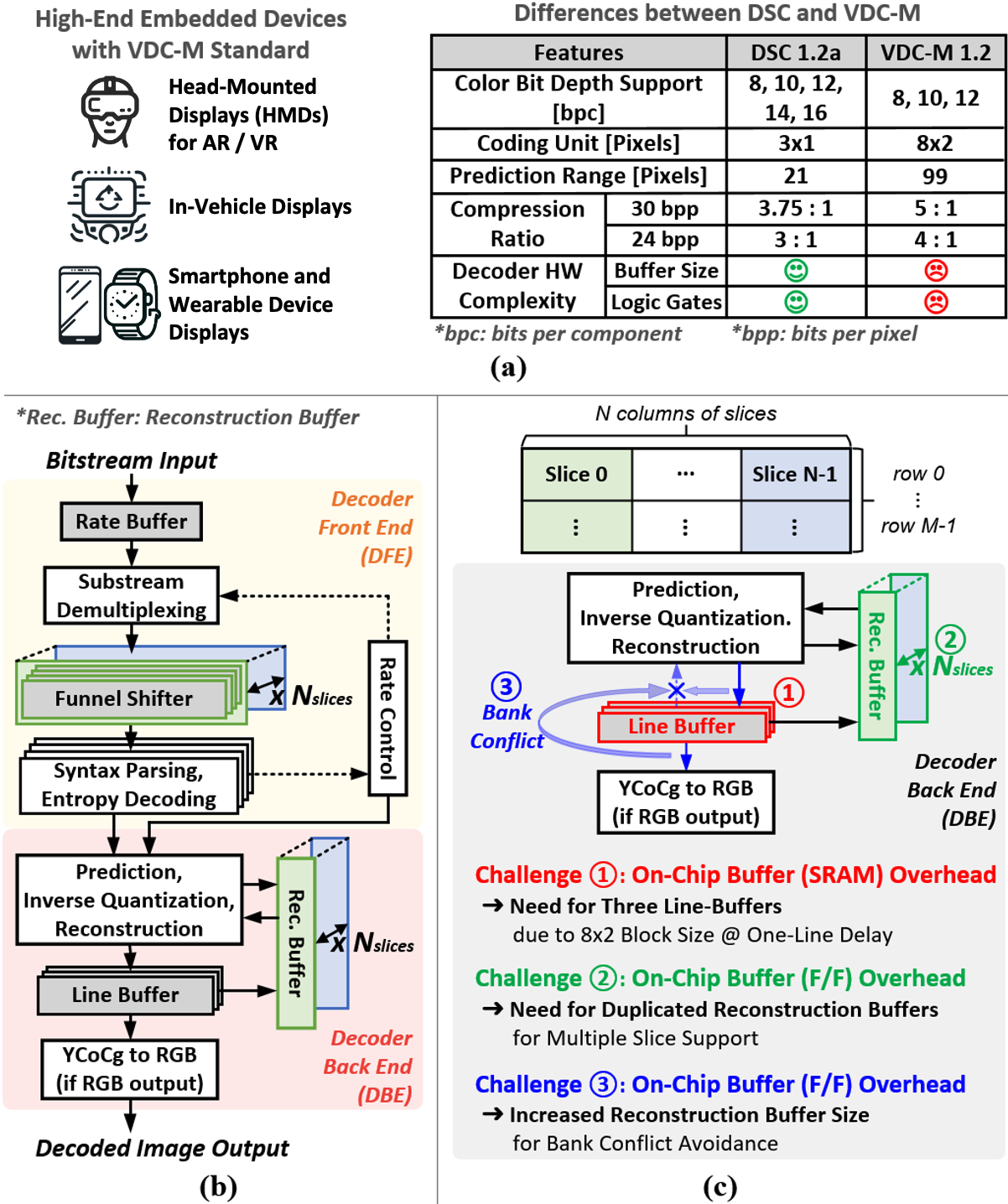}}
\caption{Overview of the VDC-M standard: (a) High-end embedded devices with VDC-M and differences between DSC, (b) a multi-slice support architecture for the VDC-M decoder, and (c) implementation challenges.}
\label{fig1}
\end{figure}

Despite the availability of various codec proposals, including H.264, H.265, H.266, JPEG, and others, the DSC series and VDC-M are widely deployed in various applications, especially for high-end embedded devices \cite{b3}. The VDC-M achieves up to a visually lossless compression ratio of 5:1, effectively reducing the data size of 30 bits per pixel (bpp) images to just 6 bpp \cite{b4}. It offers an enhanced compression ratio compared to the DSC, which supports a visually lossless compression of 3:1 \cite{b5}. This improvement stems from the expansion of the coding unit and the prediction range, as shown in Fig. 1(a). As a result, the VDC-M decoder requires a larger buffer size and more logic gates, trading off a higher compression ratio \cite{b6}.

\begin{figure*}[t]
\centerline{\includegraphics[width=18cm]{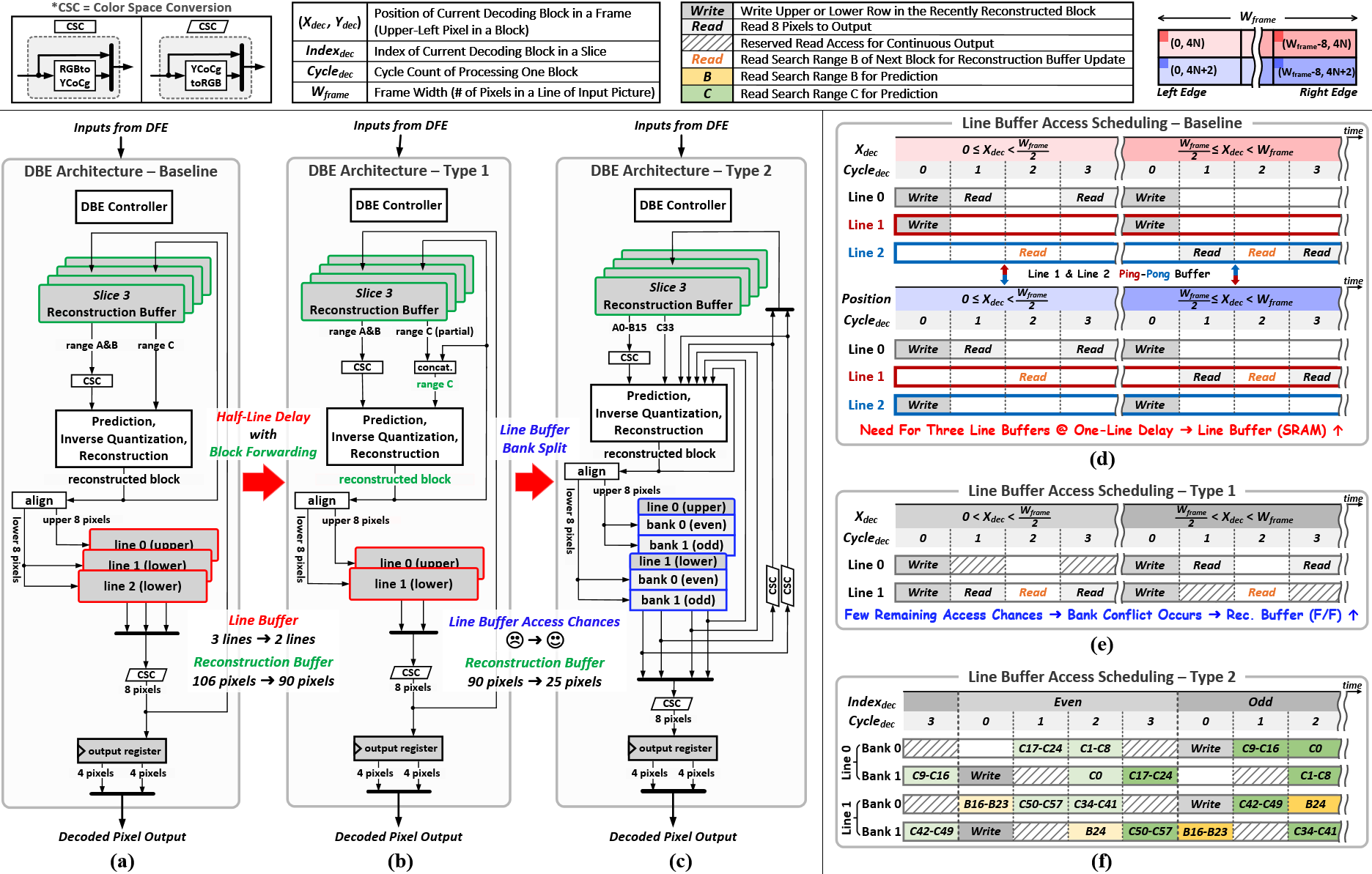}}
\caption{Proposed optimization process: DBE architecture for (a) Baseline, (b) Type 1, and (c) Type 2 with line buffer access scheduling for (d) Baseline, (e) Type 1, and (f) Type 2. Baseline architecture is the least optimized version, with optimization levels and control complexity increasing towards Type 2.}
\label{fig2}
\end{figure*}

In the VDC-M encoding process, the input image is divided into nonoverlapping, consecutive slices of identically sized rectangle. Each slice is encoded independently, enabling efficient parallel processing across multicore systems and improving error resilience by containing transmission errors within individual slices, thereby maintaining higher video quality. On the VDC-M decoder side, the funnel shifter and reconstruction buffer need to be duplicated for each slice column to enable independent processing, which increases the on-chip buffer size, as illustrated in Fig. 1(b). Therefore, it is important to minimize the associated buffer size in the design of a VDC-M decoder that supports multi-slice operation.

This paper proposes an optimization process for the memory system architecture with three architecture types, progressively increasing optimization levels and management complexity. To the best of the authors’ knowledge, this is the first published hardware implementation for the VDC-M decoder. 
\section{Motivation}
\begin{figure*}[t]
\centerline{\includegraphics[width=18cm]{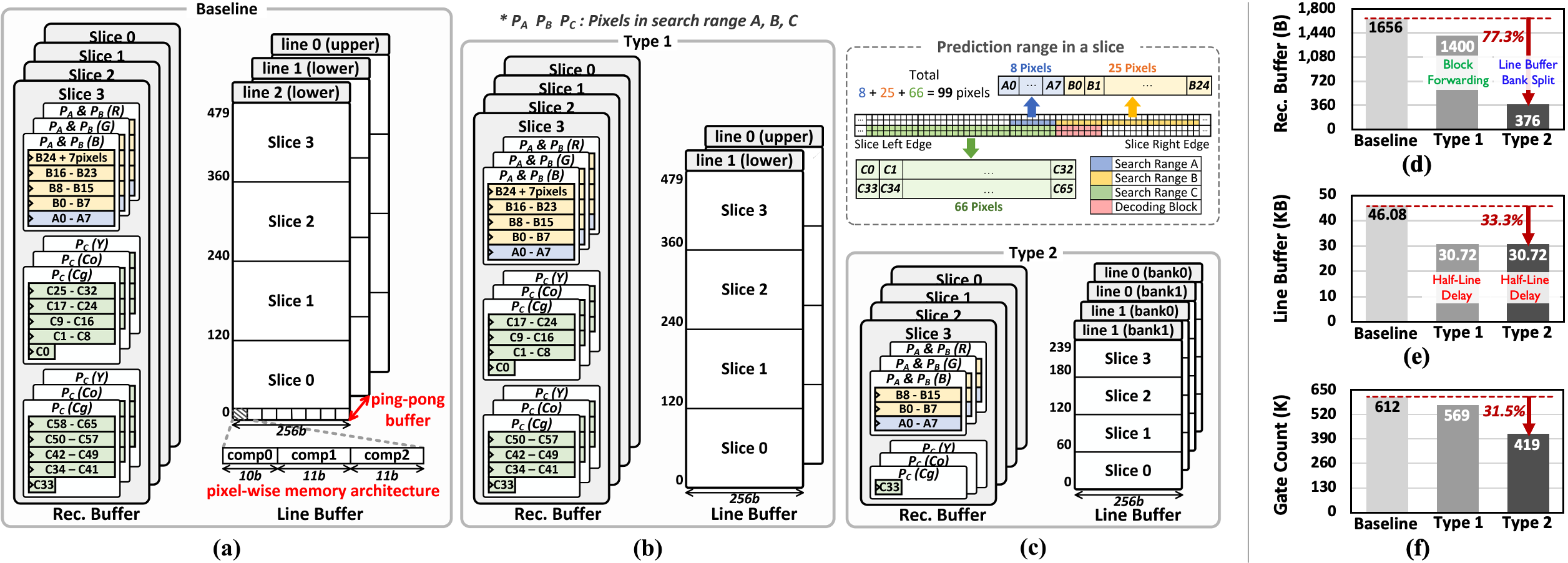}}
\caption{Proposed memory system architectures with prediction range in a slice for (a) Baseline, (b) Type 1, and (c) Type 2, and optimization result of (d) reconstruction buffer, (e) line buffer, and (f) DBE gate count.}
\label{fig3}
\end{figure*}

The extensive coding unit and prediction range of the VDC-M necessitate a larger buffer size on the decoder side, trading off a higher compression ratio. Supporting multi-slice columns exacerbates on-chip buffer overhead due to duplication of the associated buffers. Moreover, the VDC-M decoder is deployed on the display driver IC (DDI) side, where hardware resources are constrained, emphasizing the necessity for optimizing memory system architecture \cite{b7}. As shown in Fig. 1(b), the VDC-M decoder is composed of two main parts according to the characteristics of operations and on-chip buffers: the Decoder Front End (DFE) and the Decoder Back End (DBE). In the DFE, the input bitstream is stored in the rate buffer and transferred to the funnel shifter. The sizes of these two buffers are determined by standard-defined equations to prevent buffer underflow or overflow, offering a limited optimization scope. In the DBE, the line buffer stores reconstructed pixels and transfers them to output, while the reconstruction buffer holds decoded pixels for prediction. Predominantly, the prediction consumes most of the DBE's on-chip buffer, requiring roughly 100 pixels per slice column. Since there are no architectural restrictions on the line buffer and reconstruction buffer, the DBE has ample optimization opportunities. Therefore, this paper concentrates on optimizing the DBE's memory system architecture.

Fig. 1(c) shows the implementation challenges of the multi-slice support architecture, particularly on the DBE side: Challenge 1 is that with one-line delay, three line buffers are required. Line delay refers to the latency between the initial writing of decoded pixels into the line buffer and their subsequent reading out. Since each slice comprises non-overlapping 8x2 pixel blocks arranged in a 2-D array and operations are performed at the block level, decoding is simultaneously applied to two lines (i.e., a blockline), requiring at least three line buffers with one-line delay to prevent data overwriting in the lower-line buffer; Challenge 2 is reconstruction buffer overhead due to the hardware copies for supporting multi-slice operation; Challenge 3 arises from bank conflict when fetching predicted pixels directly from the line buffer to minimize the size of the reconstruction buffer;

\section{Proposed Memory System Architectures}
This section proposes three memory system architectures with progressively increasing optimization levels and management complexity. Baseline architecture evolves into Type 1 and Type 2 by applying line-delay adjustment, reconstructed block forwarding, and line buffer bank split.

\subsection{Baseline}
Baseline architecture was designed based on the reference software model \cite{b8}, incorporating necessary modifications to some buffers for practical hardware implementation. Fig. 2(a) illustrates Baseline DBE architecture, which meets system requirements supporting up to four slice columns, chroma formats of 4:4:4 and 4:2:2, and a decoding throughput of 4 pixels per cycle at an operating frequency of 200 MHz for real-time processing. It consists of three line buffers, reconstruction buffers, a controller, and functional blocks. Reconstructed data is sent to the display after the first line of the input image has been stored in the line buffer, introducing a one-line delay. To prevent data overwriting caused by the 8x2 pixel block structure, the lower line buffers operate as ping-pong buffers: one buffer outputs the previous line while the other stores the currently reconstructed block, with roles alternating after every blockline. Fig. 2(d) shows the detailed line buffer scheduling.

Fig. 3(a) shows Baseline memory system architecture. The line buffers comprise one upper and two lower lines, each storing a row in a block. These buffers utilize single-port SRAM with a pixel-wise memory architecture, featuring a 256-bit data width and 480-word lines, where 8 pixels are stored in a word. Dynamic memory allocation allows the line buffer to adjust storage based on the number of slice columns: the entire buffer space is utilized for a single slice column, while half or a quarter of the space is allocated for two or four slice columns, respectively. The reconstruction buffer is implemented with D-FF and includes an additional 7 pixels exceeding the boundaries of the prediction range B, which is necessary to provide pixels in the previous line, storing a total of 106 pixels. Four copies of the reconstruction buffer are required to support up to 4 slices per line. Each reconstruction buffer comprises three sets of circular buffers with three components. $P_A$, $P_B$, and $P_C$ mean pixels in the prediction range A, B, and C. $P_A$ and $P_B$ are stored in RGB color space while $P_C$ are stored in YCoCg color format.

\begin{figure*}[t]
\centerline{\includegraphics[width=18cm]{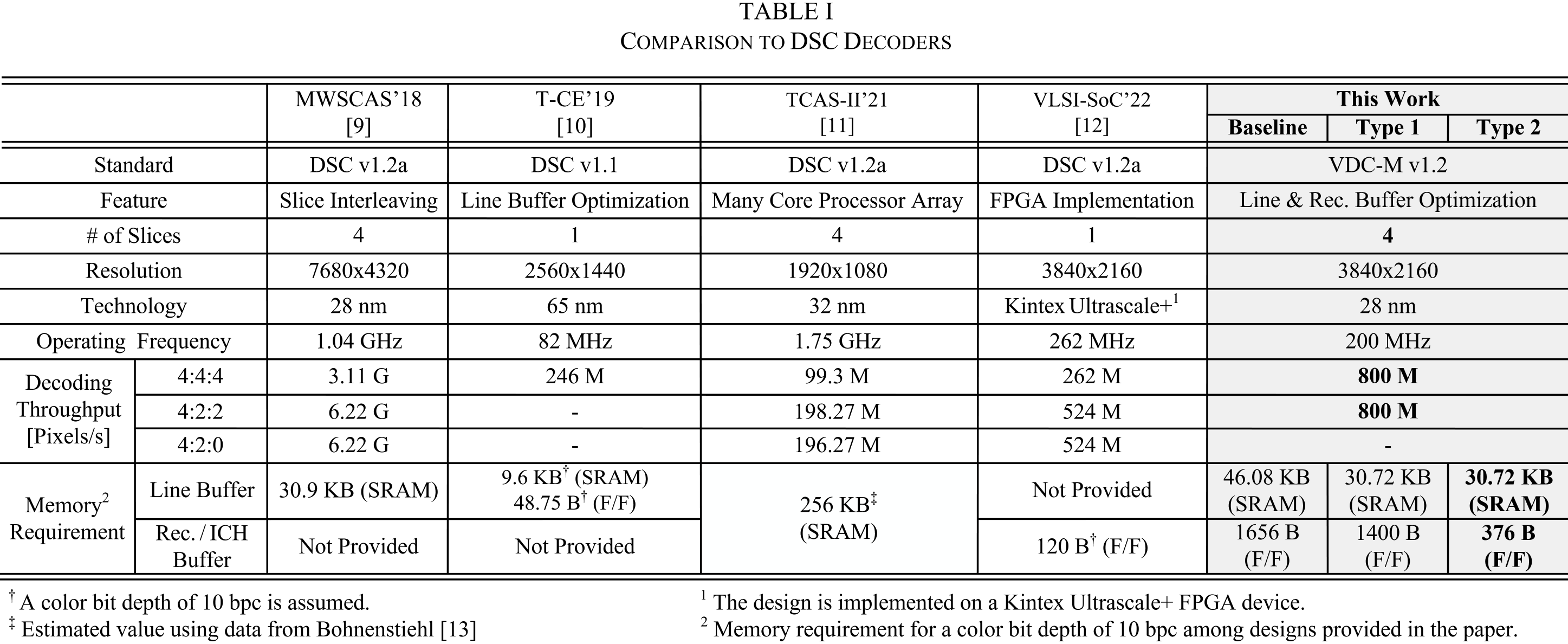}}
\label{fig10}
\end{figure*}

\subsection{Type 1: Half-Line Delay with Block Forwarding}
Fig. 2(b) illustrates Type 1 architecture, which reduces the number of line buffers to 2 lines with a half-line delay and removes 16 pixels per reconstruction buffer with block forwarding. The line buffers output restored pixels after only half of the first line is stored. Fig. 2(e) shows the line buffer access strategy: the upper line (line 0) is read out during the decoding of the first half of the line, and the lower line (line 1) is transferred for the remaining half. Notably, within the four decoding cycles, the $4K^{th}$ cycle is designated for writing the most recently restored block, and the $(4K+1)^{th}$ and $(4K+3)^{th}$ cycles are dedicated to outputting pixels in the line buffer. The $(4K+2)^{th}$ cycle is set aside for fetching predicted pixels for the next block's prediction. These reserved accesses cause bank conflict when fetching additional predicted pixels directly from the line buffer, requiring the reconstruction buffer to store nearly all pixels within the prediction range. It prevents further optimization in the reconstruction buffer.

As depicted in Fig. 3(b), the line buffer stores one upper and one lower line. The reconstruction buffer size is slightly decreased with block forwarding. Since the 16 pixels in the prediction range C25-C32 and C58-C65 are the recently restored block, they can be directly forwarded to the input for the prediction. This forwarded block is stored as C17-C24 and C50-C57 in the reconstruction buffer to decode the next block.

\subsection{Type 2: Line Buffer Bank Split}
Type 2 is the most optimized architecture. In Type 1, even though the size of the line buffer has significantly decreased, the reconstruction buffer is still burdensome. A bank split of the line buffer based on the even/odd index of the current decoding block has been employed to alleviate this overhead. It increases the line buffer access chances with the same SRAM size. As shown in Fig. 2(c), two line buffers have two banks each. The line buffer only stores the reconstructed block with an even or odd index, and the predicted pixels are directly fetched from the line buffer every cycle, as scheduled in Fig. 2(f). While line buffer management becomes complicated, it enables a reconstruction buffer only to contain 25 pixels in a slice, as depicted in Fig. 3(c).

\section{Implementation Results}
Based on the proposed DBE architectures, three VDC-M decoders were implemented with a compatible DFE architecture. The functionality was systematically verified against the official standard using the VESA Conformance Test Guideline (CTG), as documented in \cite{b8}. The proposed designs, which incorporate a buffer synthesized using flip-flop, were implemented with a 28 nm CMOS process. As depicted in Fig. 3(d), Type 2 achieves a 77.3\% size reduction in the reconstruction buffer by forwarding previously decoded block and splitting the line buffer bank. The line buffer also decreased by 33.3\%, adjusting the timing for transferring restored pixels to output from a one-line delay to a half-line delay, as shown in Fig. 3(e). Fig. 3(f) compares the hardware complexity of the proposed DBE architectures. The number of gate counts was reduced by 31.5\% compared to Baseline architecture.

Table I presents a comparative analysis of the proposed architectures with other decoders. As it is the first published work on hardware implementation for the VDC-M decoder, it mainly compares with the DSC decoders—the previously established standard used in similar mobile applications. Type 2 provides a more efficient per-slice reconstruction buffer than the DSC’s ICH buffer, supporting up to four slice columns with a total buffer size of 376 B, or 94 B per slice. This is smaller than the 120 B ICH buffer for a single slice in \cite{b12}. In contrast, except for reference \cite{b11}, the line buffer size is similar or larger despite a reduction of 16 KB. This results from the 8x2 pixel block structure of the VDC-M. Since decoding is applied to two lines simultaneously, at least two line buffers are required to prevent data overwriting in the lower line pixels. On the other hand, the DSC, with its 3x1 pixel coding unit, can function accurately with just a single line buffer. This line buffer overhead in the VDC-M is a trade-off for achieving a higher compression ratio. The larger coding unit in the VDC-M facilitates relatively higher decoding throughput by allowing more concurrent pixel processing. The proposed architectures ensure a sufficient throughput of 800 Mpixels/s, enabling real-time performance with up to 96.45 fps for 4K UHD resolution by operating at 200 MHz with a decoding rate of 4 pixels/cycle.

\section{Conclusion}
This study presents the first hardware implementation of the VDC-M decoder, proposing three architectures based on optimization levels and management complexity. The main focus was optimizing on-chip buffers, specifically the line and reconstruction buffers, which are key resource consumers. Techniques with half-line delay, bank split, and block forwarding reduced line buffer size by 33.3\% and reconstruction buffer by 77.3\%, respectively. Synthesized using a 28 nm CMOS process, the architectures achieved a 31.5\% gate count reduction in the decoder backend (DBE). Fully compatible with VDC-M v1.2, proposed architectures support up to 4K UHD resolutions at 96.45 fps, chroma formats of 4:4:4 and 4:2:2, and a maximum of 4 slices per line, operating at 200 MHz with 4 pixels/cycle throughput for real-time performance.

\section*{Acknowledgments}
This paper was the result of the research project supported by DB Kim Jun Ki Cultural Foundation. The EDA Tool was supported by the IC Design Education Center.

\end{document}